\documentclass[conference, 10 pt]{IEEEtran}  

\IEEEoverridecommandlockouts

\usepackage{titlesec}

\usepackage{cite}
\usepackage{amsmath,amssymb,amsfonts}
\usepackage{algorithmic}
\usepackage{graphicx}
\usepackage{textcomp}
\usepackage[dvipsnames]{xcolor}

\usepackage{geometry}
\usepackage{booktabs}
\usepackage{ulem}
\def\BibTeX{{\rm B\kern-.05em{\sc i\kern-.025em b}\kern-.08em
    T\kern-.1667em\lower.7ex\hbox{E}\kern-.125emX}}

\usepackage{scalerel}
\usepackage{tikz}
\usetikzlibrary{svg.path}

\definecolor{orcidlogocol}{HTML}{A6CE39}
\tikzset{
  orcidlogo/.pic={
    \fill[orcidlogocol] svg{M256,128c0,70.7-57.3,128-128,128C57.3,256,0,198.7,0,128C0,57.3,57.3,0,128,0C198.7,0,256,57.3,256,128z};
    \fill[white] svg{M86.3,186.2H70.9V79.1h15.4v48.4V186.2z}
                 svg{M108.9,79.1h41.6c39.6,0,57,28.3,57,53.6c0,27.5-21.5,53.6-56.8,53.6h-41.8V79.1z M124.3,172.4h24.5c34.9,0,42.9-26.5,42.9-39.7c0-21.5-13.7-39.7-43.7-39.7h-23.7V172.4z}
                 svg{M88.7,56.8c0,5.5-4.5,10.1-10.1,10.1c-5.6,0-10.1-4.6-10.1-10.1c0-5.6,4.5-10.1,10.1-10.1C84.2,46.7,88.7,51.3,88.7,56.8z};
  }
}

\newcommand\orcidicon[1]{\href{https://orcid.org/#1}{\mbox{\scalerel*{
\begin{tikzpicture}[yscale=-1,transform shape]
\pic{orcidlogo};
\end{tikzpicture}
}{|}}}}

\usepackage{hyperref} 

\usepackage{atbegshi}
\usepackage[absolute,overlay]{textpos}
\usepackage{rotating}
\usepackage[english]{babel}
\usepackage{blindtext}

\usepackage{framed}

\begin{document}

\title{A LiDAR-Driven Fallback Longitudinal Controller for Safer Following in Sudden Braking Scenarios}




\author{Mohamed Sabry\orcidicon{0000-0002-9721-6291} \textit{Member, IEEE}, Enrico Del Re \orcidicon{0009-0002-6417-5902} \textit{Member, IEEE}, \\ Walter Morales-Alvarez\orcidicon{0000-0001-6912-4130} \textit{Member, IEEE} and Cristina Olaverri-Monreal\orcidicon{0000-0002-5211-3598} \textit{Senior Member, IEEE}%
\thanks{Department Intelligent Transport Systems, Johannes Kepler University Linz, Altenberger Straße 69, 4040 Linz, Austria.
\texttt{
        \{mohamed.sabry, enrico.del\_re, walter.morales\_alvarez, cristina.olaverri-monreal\}@jku.at
    }
}%
}

\maketitle

\begin{abstract}
Adaptive Cruise Control has seen significant advancements, with Collaborative Adaptive Cruise Control leveraging Vehicle-to-Vehicle communication to enhance coordination and stability. However, the reliance on stable communication channels limits its reliability. Research on reducing information dependencies in Adaptive Cruise Control systems has remained limited, despite its critical role in mitigating collision risks during sudden braking scenarios. This study proposes a novel fallback longitudinal controller that relies solely on LiDAR-based distance measurements and the velocity of a follower vehicle. The controller is designed to be time-independent, ensuring operation in the presence of sensor delays or synchronization issues. Simulation results demonstrate that the proposed controller enables vehicle-following from standstill and prevents collisions during emergency braking, even under minimal onboard information.
\end{abstract}



\begin{textblock*}{18.15cm}(1.55cm,26cm) 
\begin{minipage}{17.8cm}
     \vspace{0.1cm} 
     {\footnotesize\copyright 2025 IEEE. Personal use of this material is permitted. Permission from IEEE must be obtained for all other uses, in any current or future media, including reprinting/republishing this material for advertising or promotional purposes, creating new collective works, for resale or redistribution to servers or lists, or reuse of any copyrighted component of this work in other works. DOI: 10.1109/ICVES65691.2025.11376297}
\end{minipage}
\end{textblock*}

\section{Introduction}

Adaptive Cruise Control (ACC) systems have evolved significantly in recent years to meet the growing demand for safer, more efficient driving. By continuously monitoring the speed and distance of lead vehicles, ACC systems help maintain safe following distances, enhance comfort, and reduce collision risks. These systems leverage onboard sensors and measurements for robust longitudinal control. The integration of Vehicle-to-Vehicle (V2V) communication has further enhanced ACC capabilities, particularly in vehicle platooning scenarios.

By integrating real-time data from surrounding vehicles, ACC systems can be enhanced to Collaborative Adaptive Cruise Control (CACC) \cite{ validi_2024, tu2019longitudinal}. CACC allows vehicles to anticipate traffic dynamics, improve traffic flow, and form tightly coordinated platoons with enhanced safety margins. However, this reliance on V2V communication introduces vulnerabilities, as the availability, accuracy, and synchronization of shared data are critical for the effectiveness of the CACC. In scenarios involving communication disruptions, delays, or unreliability, the stability and safety of CACC systems can be compromised, 
potentially requiring driver intervention with timely prompts \cite{olaverri2018automated} or the implementation of a failsafe alternative that relies on onboard sensors. Although significant progress has been made in ACC and CACC systems, research into using minimal onboard data has slowed down. Yet, these approaches are crucial for real-world scenarios where autonomous vehicles might lose external communication or face synchronization challenges from sensor delays or onboard computational issues \cite{bonab2024nonlinear}.


 This research gap opens the way for the need for fallback control solutions capable of operating independently with minimal onboard information, ensuring vehicles can operate and avoid collisions even in adverse conditions.

To address this gap, this paper introduces a novel Fallback Longitudinal Controller (FLC) that operates solely on minimal onboard information: distance measurements from a LiDAR sensor and the onboard speed of the vehicle. Unlike conventional systems that depend on time synchronization \cite{xu2015integrated}, the proposed FLC is designed to be temporally independent, ensuring resilience to delays and inconsistencies in sensor data or relative timestamps. This feature makes it robust to communication disruptions and synchronization errors, which are common in real-world autonomous driving scenarios. The LiDAR sensor was chosen as the primary perception sensor in this study due to its compatibility with the proposed FLC, which is designed for implementation in an ACC and steering assist system within a vehicle-following pipeline. This application demands smooth velocity transitions, a wide perception field, high-density data capable of precise object detection, and reliable operation under low-light or nighttime conditions, capabilities that LiDAR provides independently, without the need for integrating multiple sensing modalities \cite{thakur2024lidar}.

The FLC operates parallel to a baseline longitudinal control system within a vehicle-following pipeline. This pipeline is designed for use in scenarios where communication errors, delays, or synchronization issues occur, allowing the follower vehicle to maintain safe distances from the lead vehicle until normal communication or synchronization is restored. Simulations have been conducted to evaluate the performance of the controller. Results demonstrate that the fallback controller prevents collisions up to a braking threshold and maintains safe distances using minimal onboard information.

This paper is structured as follows: Section II reviews related work, Section III provides a detailed description of the proposed FLC, the vehicle following pipeline and simulation framework. Section IV presents the experimental setup followed by section V showing the simulation results. Finally, Section VI summarizes the contributions of this work and outlines potential directions for future research.


\section{Related Work}
\label{sec:RelatedWork} 
In this section, previous work on longitudinal control systems is reviewed, focusing on controller types, controller combinations, CACC, failsafe strategies, and minimal data requirement approaches. This analysis highlights the advances and challenges associated with vehicle control systems and their communication strategies.

\subsection{Controller Types}
Longitudinal control systems play a critical role in vehicle automation and safety across various configurations, including individual vehicles, leader-follower scenarios, and platoons.

The Proportional–Integral–Derivative (PID) controller is one of the most commonly used controllers due to its simplicity and applicability to real-world cases \cite{xu2015integrated}. As a model-free controller, PID does not require detailed vehicle models, making it robust in simpler environments. However, some PID versions are suboptimal in complex, dynamic conditions with rapidly changing traffic flows. To address this, adaptive PID controllers with online parameter tuning have been proposed \cite{sahputro2017design}, enhancing their responsiveness.

Model Predictive Control (MPC) has gained prominence as a more advanced alternative. MPC optimizes control outputs in real time \cite{sancar2014mpc}. For example, \cite{dai2023explicitly} employed MPC in conjunction with surrogate safety measures (SSM) to mitigate rear-end collisions by controlling the braking behavior. Although MPC often outperforms PID in certain scenarios, it necessitates a well-defined mathematical model and more computational resources.


Researchers have also explored alternative controllers, such as fuzzy logic controllers \cite{mattas2021safety}. The controller used time, two acceleration thresholds, and the relative speed of successive vehicles to ensure string stability. Another notable approach is the H-Infinity controller, which was applied in platooning scenarios \cite{li2017robust}. This controller relies on multiple vehicle model parameters for operation. A distributed version of the H-Infinity controller was further developed, using data from V2V communication and onboard sensors to maintain platoon stability. Additionally, \cite{farag2020design} proposed a simple tanh-based longitudinal controller to maintain equidistant spacing between vehicles in a platoon, offering a computationally efficient solution.

To overcome the limitations of single controllers, hybrid strategies have been developed. For example, Multiple-Model Switching (MMS) control uses multiple H-Infinity controllers and selects the one with the least error at each time step \cite{li2016multiple}.


\vspace{-5pt}

\subsection{Controllers in Collaborative Adaptive Cruise Control}


Approaches, such as those presented in \cite{chen2018robust} and \cite{zhang2024robust}, employ centralized control strategies to develop robust and flexible platooning control systems for Collaborative Automated Vehicles (CAVs). These methods formulate the platoon control problem as a Min-Max Model Predictive Control (MM-MPC) framework, integrating data from V2V communication and onboard sensors. The primary challenge addressed is the presence of communication delays, while complete loss of communication is not considered within the scope of these studies.

Failsafe mechanisms ensure system functionality during communication failures. For example, \cite{zhao2021distributed} proposed switching to a collaborative control mode using MPC and onboard sensors when communication links break. Although this approach operates with reduced efficiency, it maintains basic control functions of ACC.

Similarly, \cite{liu2021safety} introduced Safety Reinforced Cooperative Adaptive Cruise Control (SR-CACC), which detects communication failures and automatically switches to a sensor-based ACC strategy. This system relies on local sensor data for position, velocity, and acceleration measurements, ensuring continued operation despite communication disruptions.

Some studies have focused on reducing system dependencies by minimizing the number of sensors, modalities and communication data required for control. For instance, \cite{wei2019integrated} demonstrated vehicle-following using only V2V communication and radar data.

An even more simplified approach was presented by \cite{ferrara2004minimum}, where longitudinal control relied solely on radar, relative time, and follower vehicle speed. This minimalistic method enhances resilience against communication failures, making it suitable for resource-constrained environments.

\begin{table}[h!]
\renewcommand{\arraystretch}{0.9}
\centering
\caption{Side-by-side comparison of the data requirements between the ACC, CACC, and the proposed FLC.}
\label{tab:acc_cacc_comparison}
\resizebox{\columnwidth}{!}{ \begin{tabular}{cccc}
\toprule
\textbf{Used Data}                 & \textbf{ACC}        & \textbf{CACC}      & \textbf{ FLC } \\ \midrule
\textbf{Follower Speed}            & {\color{BrickRed}\checkmark}          & {\color{BrickRed}\checkmark}         & {\color{BrickRed}\checkmark}        \\ \midrule
\textbf{Calculated Leader Speed}   & {\color{BrickRed}\checkmark}          & {\color{BrickRed}\checkmark}         & {\color{ForestGreen}$\boldsymbol{-}$}                 \\ \midrule
\textbf{Actual Leader Speed}       & {\color{ForestGreen}$\boldsymbol{-}$}                 & {\color{BrickRed}\checkmark}         & {\color{ForestGreen}$\boldsymbol{-}$}                 \\ \midrule
\textbf{Calculated Leader Location}& {\color{BrickRed}\checkmark}          & {\color{BrickRed}\checkmark}         & {\color{BrickRed}\checkmark}        \\ \midrule
\textbf{Actual Leader Location}    & {\color{ForestGreen} $\boldsymbol{-}$}                 & {\color{BrickRed}\checkmark}         & {\color{ForestGreen}$\boldsymbol{-}$}                 \\ \midrule
\textbf{V2V Communication}         & {\color{ForestGreen}$\boldsymbol{-}$}                 & {\color{BrickRed}\checkmark}         & {\color{ForestGreen}$\boldsymbol{-}$}                 \\ \midrule
\textbf{Time}                      & {\color{BrickRed}\checkmark}          & {\color{BrickRed}\checkmark}         & {\color{ForestGreen}$\boldsymbol{-}$}                 \\ \midrule
\bottomrule
\end{tabular} }
\end{table}


Given the limited research on fallback longitudinal control systems to the knowledge of the authors, this paper proposes a novel FLC that operates in parallel with a PID-based longitudinal controller within a baseline vehicle-following pipeline. Table \ref{tab:acc_cacc_comparison} summarizes the data requirements of the general ACC, CACC, and the proposed FLC, highlighting its minimal data dependency and time-independence.

\begin{figure*}[htp!]
\centering
  \includegraphics[width=0.85\textwidth]{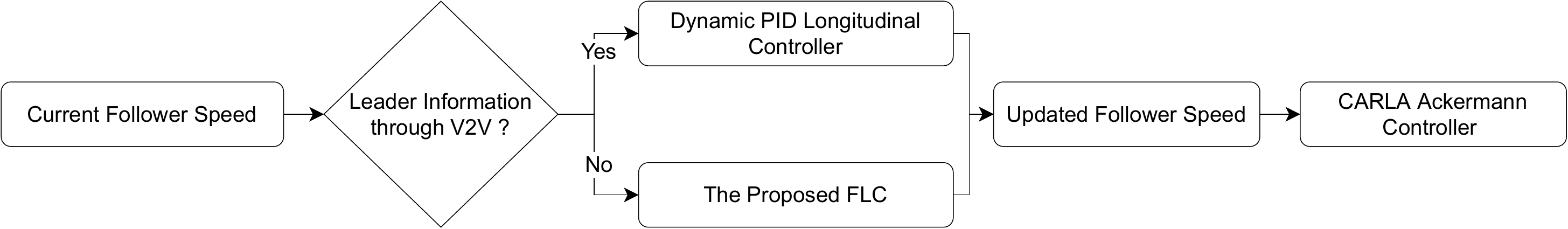}
  \caption{
  This figure illustrates the vehicle-following pipeline. The system is initialized with the current speed of the follower vehicle and checks for the availability of the lead vehicle's speed via V2V communication in the CARLA simulator. If the communication is successful and the lead vehicle's speed is successfully received, the longitudinal control is handled by a simple PID-based controller. In the absence or delay of the V2V communication data, the proposed FLC is activated to ensure continued longitudinal control.
  }
  \label{fig:follower_system}
\end{figure*}


\section{The Vehicle Follower Pipeline and the FLC} 
\label{sec:validatingSimulation}


The following section introduces the vehicle follower pipeline, which integrates the FLC within a parallel longitudinal controller module. The overall system architecture is depicted in Figure \ref{fig:follower_system}.

\subsection{Vehicle Follower Pipeline Overview} 
The vehicle follower system is designed to follow a lead vehicle starting from a stationary position, maintain consistent tracking, and avoid collisions with the lead vehicle during sudden emergency braking events, even without explicit knowledge of the lead vehicle's speed.

\subsubsection{Lead Vehicle LiDAR-Based Detector} 
To ensure realism in the testing environment, the position of the lead vehicle is determined based on a LiDAR-based obstacle detection system, as described in \cite{sabry2024autonav}. The detector divides the area in front of the follower vehicle into a 2D grid and identifies potential edge candidates. These edge candidates are processed using the Dilation Morphological operation \cite{comer1999morphological}, followed by object clustering to group detected features. The clustered objects that satisfy predefined geometric constraints are classified as vehicles. The pixel coordinates of these vehicles are reprojected into meters relative to the follower vehicle for further processing.

\subsubsection{Lateral Controller} 
Once the relative position of the lead vehicle is obtained, it is used as input for both the lateral and longitudinal controllers. For lateral control, a modified version of the pure pursuit algorithm \cite{ohta2016pure} is employed. The lateral controller takes the relative lead vehicle position, a base minimum lookahead distance, the follower vehicle's speed (in m/s), and a speed-dependent scaling factor.

The lookahead distance $D_{Ahead}$ is dynamically computed by adding the follower vehicle's speed multiplied by a scaling factor to the base lookahead distance. The Euclidean distance to the lead vehicle is then calculated, along with the relative angle between the lead and follower vehicles. The steering angle $\theta$ is computed using the element-wise arc tangent ratio between the $y$ and $x$ components of the lead vehicle's position, ensuring correct quadrant selection as per \cite{organisation1999iso}.

The resulting steering angle $\theta$ is time-independent and serves as the final input to the vehicle controller, as shown in Equation \ref{eqn:pid5}.

\vspace{-10pt}

\begin{equation}
\begin{split}
\theta = 2 * (\frac{\delta_y}{D_{Ahead}^2 + \delta_x^2})
\end{split}
\label{eqn:pid5}
\end{equation}

\subsubsection{Parallel Longitudinal Controller}


The primary longitudinal controller is a PID-based controller that assumes V2V communication with the lead vehicle. The inputs to this controller include the lead vehicle's speed, the follower vehicle's speed, the inter-vehicle distance, and the relative time gap. The controller is initialized with a desired time gap of 0.5 seconds and employs two distinct sets of proportional ($k_p$), integral ($k_i$), and derivative ($k_d$) gains to adapt its behavior based on far and near inter-vehicle distances. This configuration enhances the gap keeping capabilities of the controller.


\subsection{The Proposed FLC}

For the main contribution of this paper, the proposed instantaneous FLC is presented below. The time-independence and minimal data requirements of the controller is crucial in real-world tests, as it can operate even if a malfunction or a time synchronization issue occurs on a computational unit.

Equation \ref{eqn:pid6} shows the dynamic threshold $\tau_D$, which is an additional distance based on the follower vehicle speed $S_F$ and a fixed time gap $T_G$. $\tau_D$ is added to the base desired gap $B_{gap}$ to obtain $D_d$, which represents the distance that should be taken into consideration to keep a safe distance from the lead vehicle.

\vspace{-10pt}

\begin{equation}
\begin{split}
\tau_D = T_G * S_F \quad \\
D_d = \tau_D + B_{gap}
\end{split}
\label{eqn:pid6}
\end{equation}

The change in distance between the lead and follower vehicle in two consecutive simulation steps is used to set the threshold of emergency braking if the lead vehicle brakes abruptly $\Delta_d = (D_s - D_{s-1})$.

$Scale_f$ is a scaling factor used to get a dynamic time headway based on the follower vehicle speed $S_F$ divided by a constant $C$ (Eq. \ref{eqn:pid9}).

\vspace{-10pt}

\begin{equation}
Scale_f = min( 0.4, 0.1 * ( \frac{S_F}{C} ) )
\label{eqn:pid9}
\end{equation}

$F_{shift}$ uses the generated scaling factor $Scale_f$ to get an additional safe distance scaling with the follower vehicle speed $S_F$. The cumulative distance shift $dist_s$ is further calculated as the current distance between the leader and follower obtained from the aforementioned LiDAR-Based Detector $LF_{dist}$, with $D_d$ and $F_{shift}$ subtracted from it (Eq. \ref{eqn:pid10}). 

\vspace{-10pt}

\begin{equation}
\begin{split}
F_{shift} = max(0, S_F) * Scale_f \quad \\
dist_s = LF_{dist} - D_d - F_{shift}
\end{split}
\label{eqn:pid10}
\end{equation}

$D_d$ and $F_{shift}$ denote the desired distance between the lead and the follower vehicle, it is used for both longitudinal controllers in this paper.


\vspace{-5pt}

\begin{equation}
    a = 
\begin{cases}
    dist_s * \left( 1 - \cos\left(\pi \frac {dist_s}{bound_t}\right) \right), & \Delta_d > -0.3 \\
    dist_s * \left( 1.5 - \cos\left(\pi \frac {dist_s}{bound_t}\right) \right), & \Delta_d \leq -0.3
\end{cases}
\label{eqn:pid11}
\end{equation}

\vspace{-10pt}

\begin{figure}[h]
  \centering
  \includegraphics[width=0.8\columnwidth]{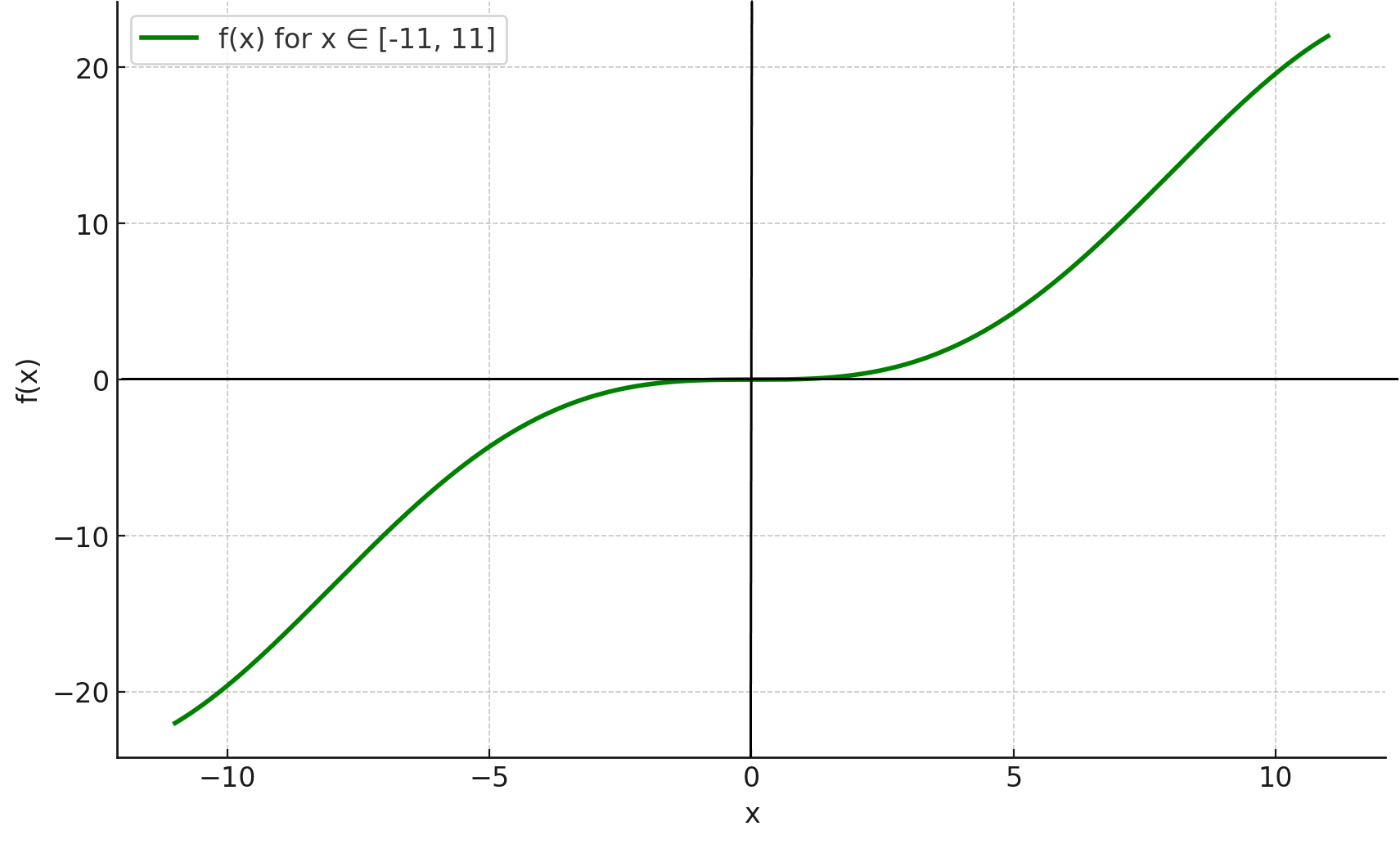}
  \caption{
  The figure illustrates the truncated cosine curve employed to regulate the acceleration of the follower vehicle when $\Delta_d$ is above -0.3. 
  } 
  \label{fig:truncated_cosine}
\end{figure}


Above is the proposed FLC time-independent controller, it is based on a truncated cosine equation (Eq. \ref{eqn:pid11}). The final $dist_s$ is used with a $bound_t$ value of 11 meters. This ensures that, at maximum boundaries the equation would get a value almost at the first peak of the cosine curve, as seen in Figure \ref{fig:truncated_cosine}. The cosine is used in this case, instead of the tanh controller for example, as it has a plateau region around the origin, which smooths the acceleration $a$ transition from positive to negative, thus reducing jerk. If the $\Delta_d$ is above -0.3 meters, the cosine equation with a shift of 1 is applied. Otherwise, a shift of 1.5 is used. This significantly reduces the plateau effect and makes the negative side of the equation more sharp, resulting in a more rapid deceleration. To further help with a more rapid deceleration, the $dist_s$ is overridden to be -3 for a fast braking action by utilizing the negative part of the custom cosine curve, seen in Figure \ref{fig:truncated_cosine}. 
To ensure safety, a multiplier further scales the negative acceleration. The instantaneous deceleration is subsequently incorporated into the current speed of the follower vehicle and input into the default longitudinal and lateral controller of the vehicle.


\subsection{ Lyapunov Stability Analysis For The Proposed FLC}

To analyze the stability and convergence of a system, a common approach involves the use of a Lyapunov function. As first introduced by \cite{lyapunov1992general}, a system described by $\dot{x} = f(x)$ is considered Lyapunov stable if there exists a positive definite function $V(x)$ with $V(x^*)=0$ for the equilibrium state $x^{*}$. For the system in Eq.~\ref{eqn:pid11}, the equilibrium state is achieved at $dist_{s}=0$, under the additional condition $S_F=S_L$, as these conditions ensure that $dLF_{dist}/dt=0$ and $dist_{s}=0$ is stable over time.

Thus, the overall system near the equilibrium point can be described as:
\begin{equation}
\begin{array}{ll}
      \frac{ dS_F }{dt} = dist_s * ( 1 - cos(\pi * \frac {dist_s}{bound_t}) ) \\
      \qquad \qquad \frac{dLF}{dt} =  S_L - S_F
\end{array} 
\end{equation}
For simplicity, Eq.\ref{eqn:pid12} is rewritten using the substitutions $x_1 = S_F - S_L$,  $x_2 = dist_s$ and $Z:=0.1/C$ as:

\begin{equation}
\begin{aligned}
    \frac{dx_1}{dt} &= x_2 \cdot \left(1 - \cos\left(\pi \cdot \frac{x_2}{bound_t}\right)\right), \\
    \frac{dx_2}{dt} &= -x_1 
    - T_G \cdot \left[x_2 \cdot \left(1 - \cos\left(\pi \cdot \frac{x_2}{bound_t}\right)\right)\right] \\
    &\quad - 2Z \cdot x_2 \cdot \left(1 - \cos\left(\pi \cdot \frac{x_2}{bound_t}\right)\right) \cdot (x_1 + S_L).
\end{aligned}
\label{eqn:pid12}
\end{equation}




Here, for $\bar{x} = (x_1,x_2)$, the equilibrium point is located at the origin. The Lyapunov function is defined as the total utilized energy of the system, as expressed in Eq.~\ref{eqn:pid13}.

\begin{equation}
\begin{array}{ll}
      V(\bar{x}) = \frac{1}{2}(x_1^2+x_2^2)
\end{array} 
\label{eqn:pid13}
\end{equation}

The function is positive definite outside of $x^*$, making it a potential candidate for proving stability around the origin. However, near $\bar{x} = (0,0)$, its derivative is not negative semi-definite. The dominant term in this region is expressed in Eq.~\ref{eqn:pid14}, obtained after approximating the cosine function:

\begin{equation}
\begin{array}{ll}
      \frac{dV(\bar{x})}{dt}  \propto -x_1x_2 + O(x^3)
\end{array} 
\label{eqn:pid14}
\end{equation}


This indicates that the system achieves stability under the following conditions:
\begin{itemize}
    \item The follower vehicle is travelling faster than the lead vehicle and is positioned farther away than the desired distance.
    \item The follower vehicle is travelling slower than the lead vehicle and is positioned closer than the desired distance.
\end{itemize}

Despite the control system being only partially Lyapunov stable due to the limited information utilized by the proposed FLC, the system will ultimately converge to the stable region, defined as $x1,x2>0$, regardless of the initial conditions.


\section{ Experimental Setup }
\label{sec:ExperimentalResults}



To evaluate the performance of the FLC, we relied on the 3DCoAutoSim \cite{olaverri2018connection, hussein20183dcoautosim, michaeler20173d} framework and CARLA. In the simulated environment, a lead vehicle and a follower vehicle were placed 7 meters apart to initiate the scenario.
The lead vehicle accelerated from a standstill and covered approximately 650 meters before performing an abrupt deceleration maneuver. Simultaneously, the rear vehicle followed the lead vehicle, adjusting its speed and maintaining a safe distance to avoid a collision during the sudden deceleration. The experiment was conducted at four predetermined maximum speeds: 30 km/h, 50 km/h, 70 km/h, and 90 km/h. For the FLC parameters, $T_G$ is set to 1.0, $B_{gap}$ is set to 6.0 and $C$ is set to $\frac{5}{3.6}$.

\begin{figure}[!ht]
  \centering
  \includegraphics[width=0.75\columnwidth]{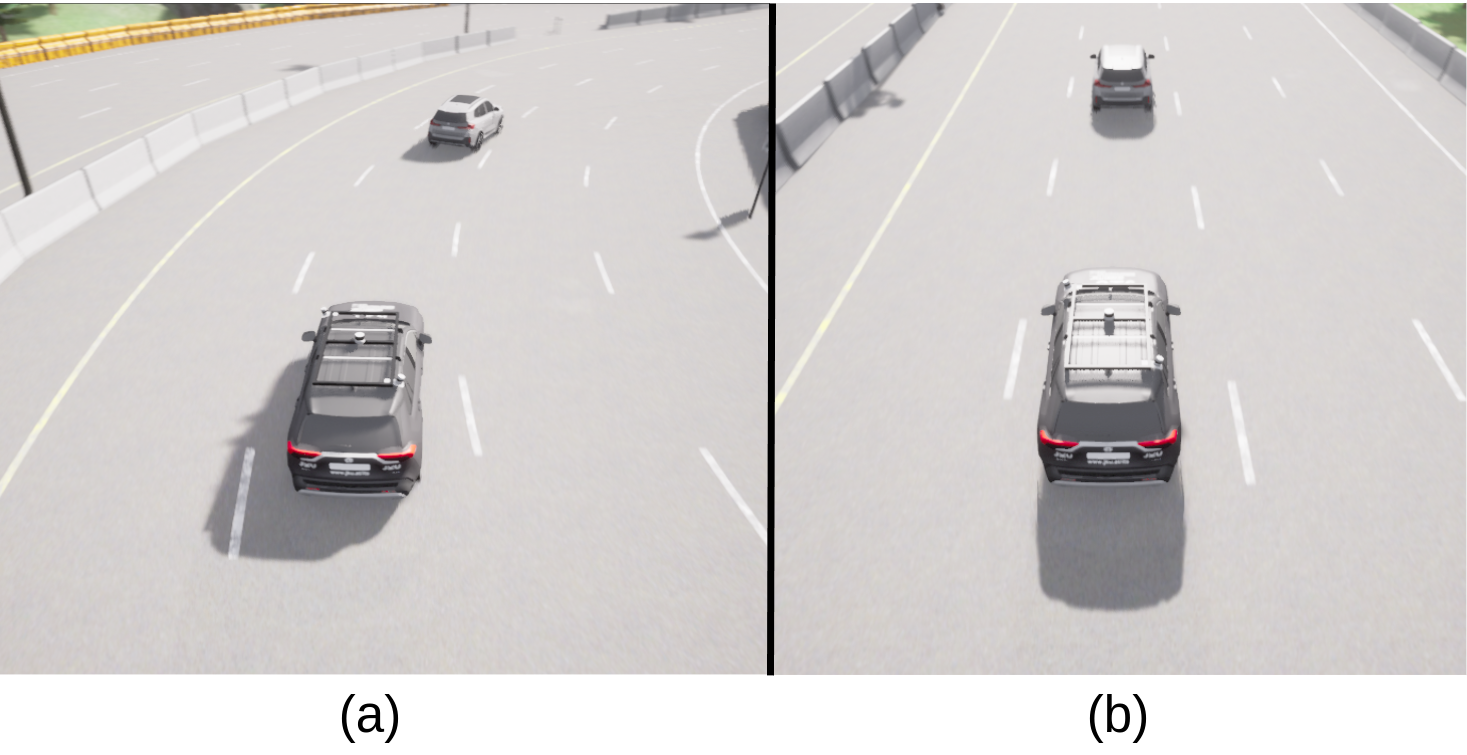}
  \caption{
  The figure presents the 3D model of the JKU-ITS research vehicle within the CARLA simulator following a lead vehicle at a speed of 70 km/h. Subfigure (a) depicts the follower vehicle tracking the lead vehicle in a curve, while subfigure (b) illustrates a simplified straight-line scenario.
  } 
  \label{fig:jku_car}
\end{figure}

Figure \ref{fig:jku_car} shows a screenshot from the scenario. 
The follower vehicle utilized only the proposed FLC to mitigate the risk of collision. During the tests, the maximum deceleration of the lead vehicle was -6 m/$s^2$, which was used in previous research to denote emergency braking scenarios \cite{svard2017quantitative}. For comparison, the same test was performed using a PID-based controller, as seen in Figure \ref{fig:pathErrors}.

\section{ Results }

\begin{figure*}  
\vspace*{-18mm}
\centering
\includegraphics[width=0.80\textwidth]{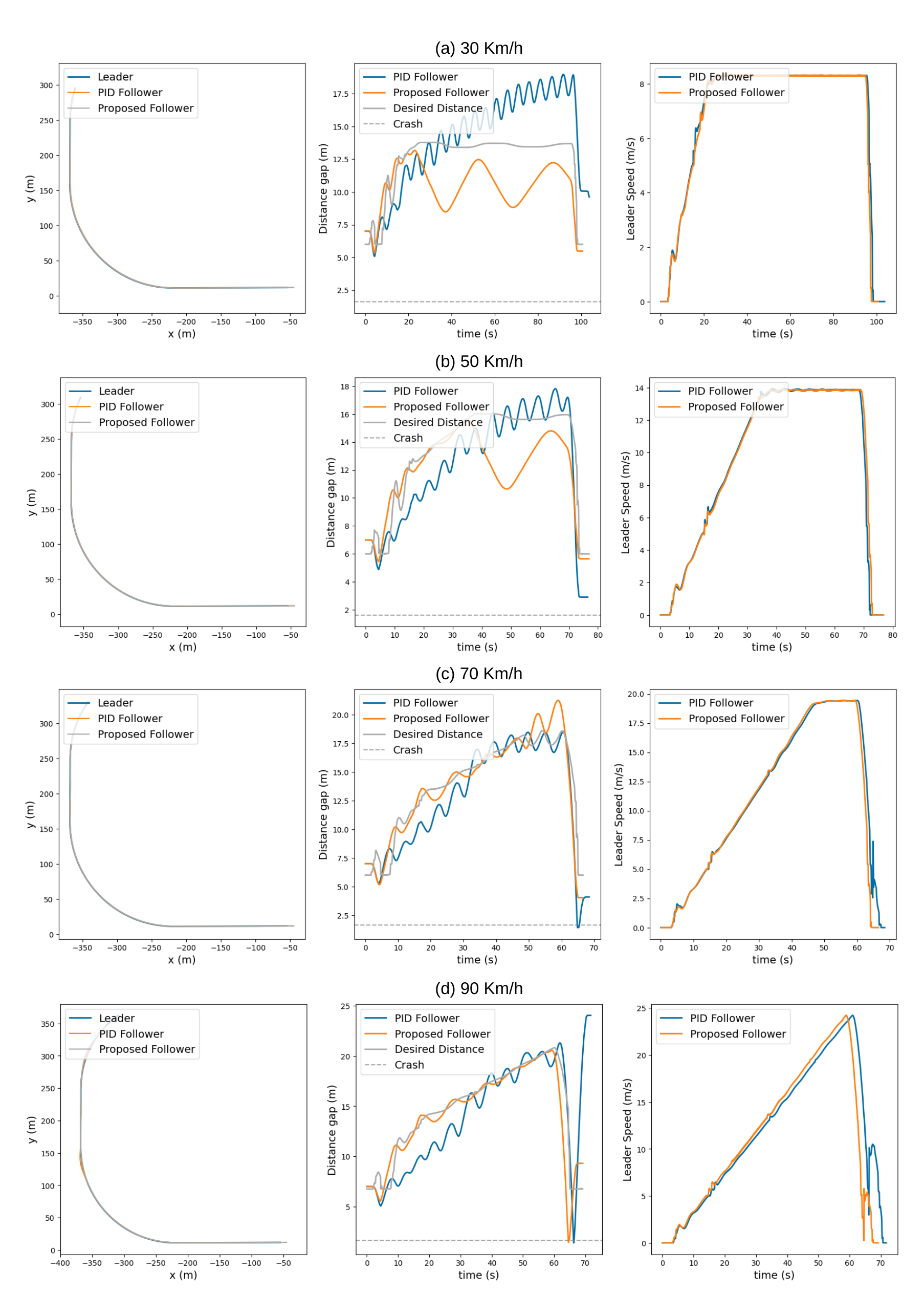}
  \caption{
  The figure shows the results from the test path in the CARLA simulator with four speeds. Over a test distance of approximately 650 meters, the lead vehicle accelerated to the target speed and traversed the aforementioned distance before performing an abrupt braking maneuver with a constant deceleration of -6 m/$s^2$. The results demonstrate that the proposed FLC successfully executed timely braking to prevent collisions at speeds up to 70 km/h. In contrast, the conventional PID controller was only capable of preventing collisions up to 50 km/h. The results are presented across three graphs for each speed: the leftmost graph overlays the trajectories of the lead vehicle and the follower vehicle controlled by both the fallback and PID controllers; the middle graph illustrates the variation in gap distance between the lead and follower vehicles for each controller in separate test runs; and the rightmost graph displays the speed profile of the lead vehicle during both runs. At 90 km/h, both controllers were unable to successfully stop the follower vehicle.
  } 
  \label{fig:pathErrors}
\end{figure*}



Figure \ref{fig:pathErrors} presents the results of the experiments, including the trajectories of the lead and follower vehicles, the speed profile of the lead vehicle, and the gap distance between the two vehicles at four different speeds. Despite the lead vehicle undergoing a sudden deceleration of -6m/$s^2$, the follower vehicle equipped with the FLC successfully stopped in the tests conducted at 30, 50 and 70 Km/h preventing a potential collision with a clearance of over 3.8, 4 and 2.3 meters respectively from the lead vehicle. The FLC operates in real time at 20 fps, requiring no V2V communication or speed information from the lead vehicle. 


The PID-based controller was successful at preventing collisions in the 30 km/h and 50 km/h tests with a clearance of over 7.9 and 1.2 meters respectively from the lead vehicle.
Both controllers failed to prevent a collision at 90 km/h; however, the proposed FLC was able to decelerate the following vehicle more than the PID, resulting in a less severe crash.

The follower vehicle using the FLC, achieved average rotational and translational errors of 0.0176° and 0.2258 meters, while the PID controller demonstrated slightly higher average rotational and translational errors of 0.0199° and 0.2876 meters respectively. The fallback controller successfully avoided a collision when the lead vehicle applied a sudden braking maneuver with a deceleration of -6m/$s^2$ up to a speed of 70 km/h, which highlights the capability of the proposed FLC to transition from a stationary state, track a lead vehicle, and promptly react to abrupt braking scenarios. Although the PID controller successfully tracked the lead vehicle under normal conditions, it failed to respond effectively above 50 km/h during the abrupt braking scenario, resulting in a collision.
At 90 km/h, both controllers were not able to handle the sudden braking maneuver.

\section{Conclusion and Future Work}
\label{sec:Conclusions}





The proposed FLC demonstrated the ability to effectively follow a lead vehicle and prevent collisions during sudden braking maneuvers up to a speed of 70 km/h. Utilizing minimal onboard data while being time-independent.
The controller adds an extra safety layer between conventional ACC and emergency braking, creating a four-stage control hierarchy when CACC is included. Integrated into a vehicle-following pipeline and relying solely on onboard sensor data, excluding radar and relative timing, it demonstrated stability under the specified braking limits and speed conditions, highlighting its potential as a backup emergency system if V2V communication, radar inputs, or computational synchronization fail. Simulation results showed stable platooning behavior, though further mathematical analysis is needed to validate convergence and stability, particularly because the FLC operates with reduced input, omitting the lead vehicle’s speed.

Future research will focus on enhancing the resilience of the fallback controller to prevent crashes at higher speeds and integrating it into more complex and challenging scenarios.

\section*{Acknowledgment}

This work was funded by the Austrian Research Promotion Agency (FFG) project pDrive number: 901692.





\bibliographystyle{IEEEtran}
\bibliography{paper}

\end{document}